\documentclass[preprint,12pt]{elsarticle}
\usepackage{amssymb}
\usepackage{graphicx}
\usepackage{hyperref}
\usepackage{color}
\journal{Chemical Physics Letters}

\begin{document}
\begin{frontmatter}
\title{Effect of bath temperature on the quantum decoherence}
\author[sjtu]{Peihao Huang}
\author[sjtu]{Hang Zheng}
\address[sjtu]{Key Laboratory of Artificial Structures and Quantum Control (Ministry of Education), Department of Physics, Shanghai Jiao Tong University, Shanghai 200240, China}

\begin{abstract}
The dynamics of a qubit in two different
environments are investigated theoretically.
The first environment is a two level system coupled to a bosonic
bath. And the second one is a damped harmonic oscillator. Based on a unitary transformation, we find that the decoherence of the qubit can be reduced with increasing temperature $T$ in the first case, which agree with the results in [Phys. Rev. Lett. 100, 120401], whereas, it can not be reduced with $T$ in the second case. In both cases, the qubit dynamics are changed substantially as the coupling increases or finite detuning appears.
\end{abstract}


\end{frontmatter}

\section{Introduction}

Quantum computation reveals advantages over classical one for its highly
efficient parallel calculation and therefore attracts wide interests among
scientists \cite{Nielsen2000}. With the development of nano-technology, solid state qubits, such as quantum dots as well as Josephson junction devices, are designed to
fulfill the scalability and avoid severe decoherence \cite{Nakamura1999,Chiorescu2003,You2005}. There is something in common in these promising
designs, that is the qubit is coupled to a
single-mode quantum structure which in turn couple to a multi-mode bath. For
example, the Josephson qubit suffers the intrinsic slow noise caused by the
two level fluctuators (TLFs) \cite{Simmonds2004}. The flux qubit is usually
coupled to a read-out device which can be viewed as a damped harmonic
oscillator (DHO) \cite{Faoro2004,Rebentrost2009}. People commonly believe
that temperature only plays a negative role in preserving the qubit
coherence. However, it is pointed out in Ref. \cite{Montina2008}
that the temperature can help the coherence when the qubit is coupled to a TLF (or spin-boson) environment. In this paper, we use a unitary transformation to deal with this problem, and find that the results agree with Ref. \cite{Montina2008}'s results when qubit-TL coupling is less than TL-boson coupling, that is $g_0<\alpha$ ($g_0$ and $\alpha$ are defined below). When $g_0>\alpha$, the decoherence is only increasing with increasing temperature. On the other hand, we also investigate the qubit dynamics under another environment, where the TL is replaced by a harmonic oscillator (HO). And in this case, we find that the decoherence can only be enhanced with increasing temperature. In both cases, the qubit shows beating dynamics when TL-boson (or HO-boson) coupling is small for the on-resonance case, and it is totally suppressed and becomes simple oscillation for large coupling or finite detuning.

\subsection{1st model: qubit coupled to a TLF}

1/f noise is prevailing in the Josephson qubits, which is probably due to
the intrinsic TLFs caused by defects or impurities \cite%
{Simmonds2004,Paladino2002,Falci2005,Shnirman2005,Abel2008,Neeley2008,Lupascu2009,Lisenfeld2010}%
. In this paper, we study the effect of a single TLF
environment which is described by the spin boson model (SBM) \cite{Weiss1999}. The
Hamiltonian of such a system reads $H=H_{A}+H_{B}+V$ with ($\hslash =1$) \cite{Lupascu2009,Lisenfeld2010,Gassmann2002,Gassmann2005,Paladino2006,Paladino2008,Oxtoby2009}
,
\begin{eqnarray}
H_{A} &=&\frac{\Delta _{A}}{2}\sigma
_{x}^{A},\,\,\,\,\,\,\,\,V=g_{0}\sigma _{z}^{A}\sigma _{z}^{B}, \\
H_{B} &=&\frac{\Delta _{B}}{2}\sigma _{x}^{B}+\sum_{k}{\omega _{k}}%
b_{k}^{\dagger }b_{k}+\frac{\sigma _{z}^{B}}{2}\sum_{k}g_{k}(b_{k}^{\dagger
}+b_{k}),
\end{eqnarray}%
where $\sigma _{x}$ and $\sigma _{z}$ are the usual pauli matrices, $b_{k}$ ($b_{k}^{\dag })$ are the
annihilation (creation) operators of the bath mode. $\Delta_A$ ($\Delta_B$) is the gap of the qubit (TL). Only transverse coupling
are included for simplicity \cite{Lupascu2009}. The bath is fully defined by
the spectral density,
\begin{equation}
J(\omega )\equiv \sum_{k}g _{k}^{2}{\delta }(\omega -\omega _{k}).
\end{equation}

Most of the studies on this model focus on the qubit dynamics without
considering the temperature effect \cite{Gassmann2002,Gassmann2005,Paladino2006,Paladino2008,Oxtoby2009}.
Recently, A. Montina et al. studied the temperature
effect in the resonance case $\Delta _{A}=\Delta _{B}$ by using Lindblad
master equation \cite{Montina2008}. They pointed out that the temperature
helps the qubit coherence. {Since the Lindblad equation is based on the
rotating wave approximation (RWA) and the Markov approximation, it requires $\gamma_B\ll\Delta_B$ to justify the RWA and $g_0\ll\gamma_B$ to justify the Markov approximation ($\gamma_B$ is the width of the bath spectrum that the system A sees, which is given by $\gamma(\omega_B)$ in Eq.~(\ref{G1approx},\ref{G2approx})).} In
this paper, based on a unitary transformation, we explore the dynamics in a much looser condition: $\Delta _{A}$ is
not necessarily equal to $\Delta _{B}$, {and $g_0$ can be bigger than $\gamma_B$}.

\subsection{2nd model: qubit coupled to a DHO}

In the last decade, many promising qubit schemes have been proposed and
realized, some of which fit our second model, such as a flux-qubit read
out by a dc-SQUID \cite{Chiorescu2003,Thorwart2004} or a qubit
placed in a leaky cavity \cite{Thorwart2000}. The Hamiltonian of the second
model is similar to the first one, only the TL is replaced by a HO, and the corresponding $V$ and $H_{B}$ are
\begin{equation}
V=g_{0}\sigma _{z}^{A}(B^{\dagger }+B),
\end{equation}
\begin{equation}
H_{B} ={\Omega }B^{\dagger
}B+\sum_{k}\omega _{k}b_{k}^{\dagger }b_{k}+(B^{\dagger }+B)\sum_{k}g _{k}(b_{k}^{\dagger }+b_{k})
+(B^{\dagger }+B)^{2}\sum_{k}\frac{g _{k}^{2}}{\omega _{k}},
\end{equation}
where $\Omega $ is the frequency of HO, $B$ (or $
B^{\dag }$) is the annihilation (or
creation) operators of HO. The last term in $H_{B}$ is the counter-term, which cancels the additional
contribution due to the coupling of the HO to the
bath\cite{Weiss1999,Costa2000}. Similar to the 1st model, the bath is fully defined by $J(\omega)$.

This model can be mapped to the SBM with a Lorentzian structured spectral density
\cite{Garg1985,Tian2002}. Since the spectral poses challenge to many existing method, it arouse much attention recently. Till now, it has been studied by many different methods including the quasi-adiabatic propagator path
integral (QUAPI) \cite{Thorwart2004,Liang2007,Liang2008}, the Van Vleck
perturbation theory together with a Born-Markov master equation \cite%
{Hausinger2008}, the flow equation renormalization \cite%
{Kleff2003,Wilhelm2004,Kleff2004}, the non-interacting blip approximation
(NIBA) \cite{Wilhelm2004,Nesi2007}, and generalized polaron transformation
method \cite{Brito2008,Huang2008,Gan2010}. Again, most of the works have not
considered the effect of temperature.

\section{Unified Treatment}

The aforementioned two models can be expressed as $H=H_{0}+V$,
\begin{equation}
H_{0}=H_{A}+H_{B},~~~~~V=g_{0}\sigma_{z}^{A}Q^{B}.
\end{equation}%
where $Q^{B}$ is $\sigma _{z}^{B}$ in the first case, $(B^{\dagger }+B)$ in
the second case. One can use the weak-coupling approximation \cite{Gassmann2002,Gardiner2000} (or the so called {rigorous Born
approximation} \cite{DiVincenzo2005,Burkard2009}) to deal with this Hamiltonian. Compared to the usual
Born-Markov approximation (Redfield equation), this method keeps the full
information contained in the correlation functions at the price of
introducing a kernel for the master equation which is no longer local in
time. Since the structured environments may have considerable memory time,
the Markov approximation may not be valid.

The total density matrix (system+environment) $\chi (t)$ obeys the
Liouville-von-Neumann equation,
\begin{equation}
\frac{d}{dt}\tilde{\chi}(t)=-i[\tilde{V}(t),\tilde{\chi}(t)],
\end{equation}%
where the tildes denote operators in the interaction picture with respect to
$H_{0}$. Iterating up to the second order and tracing out the
environmental degrees, one get the master equation within the Born
approximation,
\begin{equation}
\frac{d}{dt}\tilde{\rho}(t)=-i\mathrm{Tr}_{B}[\tilde{V}(t),\rho _{B}\otimes
\tilde{\rho}(0)]-\int_{0}^{t}dt^{\prime }\mathrm{Tr}_{B}[\tilde{V}(t),[\tilde{V%
}(t^{\prime }),\rho _{B}\otimes \tilde{\rho}(t^{\prime })]],
\label{MasterEquation1}
\end{equation}%
where $\tilde{\rho}$ is the reduced density matrix $\tilde{\rho}(t)=$Tr$_{B}%
\left[ \tilde{\chi}(t)\right] $, and $\tilde{\chi}(t)$ is replaced by an
approximate factorized density matrix $\tilde{\chi}(t)\approx \rho
_{B}\otimes \tilde{\rho}(t)$. The environment is usually assumed to remain in
thermal equilibrium $\rho _{B}={e^{-\beta H_{B}}}/{\mathrm{Tr}e^{-\beta
H_{B}}}$, which is justified when the environment is 'very large' and the coupling $%
H_{SB}$ 'weak' ($g_{0}\ll \Delta _{A},\Delta _{B},\Omega )$, so that the
back-action of the system onto the environment can be neglected. Going back to the
schr\"{o}dinger picture and inserting $V=g_{0}\sigma _{z}^{A}Q^{B}$, we get%
\begin{equation}
\frac{\partial \rho _{A}(t)}{{\partial }t}=-i\left[ H_{A},\rho _{A}(t)\right]
-\int_{0}^{t}d\,t^{\prime }\,X(t,t^{\prime })  \label{ME_S}
\end{equation}%
with
\begin{eqnarray*}
X(t,t^{\prime }) &\equiv &g_{0}^{2}G_{1}({t^{\prime }})\sigma
_{z}^{A}e^{-iH_{A}t{^{\prime }}}\sigma _{z}^{A}\rho _{A}(t{-t^{\prime }}%
)e^{iH_{A}t{^{\prime }}} \\
&-&g_{0}^{2}G_{1}({t^{\prime }})e^{-iH_{A}t{^{\prime }}}\sigma _{z}^{A}\rho
_{A}(t{-t^{\prime }})e^{iH_{A}t{^{\prime }}}\sigma _{z}^{A} \\
&+&g_{0}^{2}G_{2}({t^{\prime }})e^{-iH_{A}t{^{\prime }}}\rho _{A}(t{%
-t^{\prime }})\sigma _{z}^{A}e^{iH_{A}t{^{\prime }}}\sigma _{z}^{A} \\
&-&g_{0}^{2}G_{2}({t^{\prime }})\sigma _{z}^{A}e^{-iH_{A}t{^{\prime }}}\rho
_{A}(t{-t^{\prime }})\sigma _{z}^{A}e^{iH_{A}t{^{\prime }}},
\end{eqnarray*}%
where, $G_{1}({t})$ and $G_{2}({t})$ are the correlation functions $\langle
Q^{B}{(t)}Q^{B}\rangle _{\beta }$ and $\langle Q^{B}Q^{B}{(t)}\rangle
_{\beta }$, in which $\langle \,\cdots \rangle _{\beta }$ represents the
average with thermodynamic probability $\rho _{B}$. Therefore, $G_{1}({t})$
and $G_{2}({t})$ contain all the information of the structured environment.

The master equation Eq.~(\ref{ME_S}) which is a $2\times 2$ matrix equation
can be solved exactly by the Laplace transform since the convolution theorem
can be applied to the equation of each matrix element. Here, for simplicity, we only present the comparatively brief expression.
Suppose the system is in the upper eigenstate of $\sigma _{z}$ at time $t=0$, the population difference $P(t)\equiv \langle \sigma_{z}^{A}(t)\rangle \equiv \mathrm{Tr}_{A}(\sigma _{z}^{A}\rho _{A}(t))$ can be obtained in the Laplace space as
\begin{equation}\label{Ps}
\overline{P(s)}=\frac{s+2F(s)}{s^{2}+2sF(s)+\Delta _{A}^{2}},
\end{equation}%
where
\begin{equation}\label{Fs}
F(s)\equiv \int_{-\infty }^{\infty }d\omega \,J_{eff}(\omega )/(s+i{%
\omega }),
\end{equation}
\begin{equation}\label{Jeff}
J_{eff}(\omega )=g_{0}^{2}\left(G_{1}({\omega })+G_{2}({\omega })\right).
\end{equation}

\section{Correlation functions}

\subsection{1st Case: TLF Environment}

From the above derivation, we can see that, the dynamics of the qubit is
completely determined by the correlation functions $\langle Q^{B}{(t)}%
Q^{B}\rangle _{\beta }$ and $\langle Q^{B}Q^{B}{(t)}\rangle _{\beta }$
within the weak-coupling approximation. To obtain the correlation functions
more accurately, we apply a unitary transformation to the Hamiltonian, $%
H^{\prime }=\exp (S)H\exp (-S)$, with the generator $S\equiv \sum_{k}\frac{g_{k}}{2\omega _{k}}\xi _{k}(b_{k}^{\dag
}-b_{k})\sigma _{z}^{B}$. The purpose is to transform to a better representation
in which the exact solvable term contains the most important physics. Similar to Ref.~\cite{Huang2008,Zheng2004}, with the choice of
\begin{equation}\label{xi}
\xi _{k}=\frac{\omega _{k}}{\omega _{k}+\eta \Delta _{B}},
\end{equation}
\begin{equation}\label{eta}
\eta =\exp \left[ -\sum\limits_{k}\frac{g_{k}^{2}}{2\omega _{k}^{2}}\xi
_{k}^{2}\coth (\beta \omega _{k}/2)\right],
\end{equation}
where, finite temperature is considered in $\eta$, $H_{B}$ is transformed to $H_{B}^{\prime }=H_{0}^{\prime
}+H_{1}^{\prime }$ ($H_{A}$ and $V$ are not affected) with
\begin{eqnarray}
H_{0}^{\prime } &=&\frac{\eta \Delta _{B}}{2}\sigma _{x}^{B}+\sum\limits_{k}{%
\omega _{k}}b_{k}^{\dagger }b_{k},  \label{H_0'} \\
H_{1}^{\prime } &=&\sum_{k}V_{k}(b_{k}^{\dagger }\sigma _{-}^{B}+b_{k}\sigma
_{+}^{B}),
\end{eqnarray}%
where, $V_{k}={\ g_{k}\eta \Delta }_{B}/{\ (\omega _{k}+\eta \Delta _{B})}$
and $\sigma _{\pm }^{B}\equiv (\sigma _{z}^{B}{\mp }i\sigma _{y}^{B})/2$. A trivial constant and the terms of the order of $g_{k}^2$ and higher, have been omitted \cite{Huang2008,Zheng2004}.
One can see that the renormalized TL-bath coupling $V_{k}<g_{k}$, which enables the subsequent
second order perturbation (Born approximation) well conditioned compared
with the direct perturbation to the original Hamiltonian.

The equation of motion of Green's function reads,
\begin{equation}
\omega \langle \langle A|\,B\rangle \rangle _{\omega }^{\prime }=\langle
[A,B]_{+}\rangle ^{\prime }+\langle \langle [A,H_{B}^{\prime
}]|\,B\rangle \rangle _{\omega }^{\prime },
\end{equation}%
where $\langle \langle A|\,B\rangle \rangle ^{\prime }$ represents the
Fourier transform of the Green's function $-i\theta (t)\langle [A,B]_{+}\rangle _{\beta }^{\prime }$, ($[,]_{+}$ is the anti-commutator and $%
[,]$ the commutator). If we substitute $b_{k}^{\dagger }b_{k}$ by its thermodynamic
average value $n_{k}$ and omitted all the $b_{k}^{\dagger }b_{k}^{\dagger }$
and $b_{k}b_{k}$ terms, the equation chain will be
self-closed, and we get
\begin{equation}
{\langle \langle }\sigma _{z}^{B}|\,\sigma _{z}^{B}\rangle \rangle _{\omega
}^{\prime }=\frac{1}{\omega -\eta \Delta -\sum_{k}\frac{V_{k}^{2}(2n_{k}+1)}{%
\omega -\omega _{k}}}+\frac{1}{\omega +\eta \Delta -\sum_{k}\frac{%
V_{k}^{2}(2n_{k}+1)}{\omega +\omega _{k}}}.
\end{equation}%
Since ${\sigma _{z}^{B}}$ commutate with the genertator $S$, we expect ${%
\langle \langle }\sigma _{z}^{B}|\,\sigma _{z}^{B}\rangle \rangle _{\omega }=%
{\langle \langle }\sigma _{z}^{B}|\,\sigma _{z}^{B}\rangle \rangle _{\omega
}^{\prime }$. Consequently, according to the fluctuation dissipation theorem
(FDT), the correlation functions $G_{1}({t})$ and $G_{2}({t})$ can be
expressed by the retarded Green's function $\langle \langle \sigma
_{z}^{B}|\,\sigma _{z}^{B}\rangle \rangle _{\omega +i0^{+}}^{\prime }$. In the Fourier space,
\begin{eqnarray}
G_{1}({\omega }) &=&-1/\pi \mathrm{Im}\langle \langle \sigma
_{z}^{B}|\,\sigma _{z}^{B}\rangle \rangle _{\omega +i0^{+}}^{\prime
}/(1+e^{-\beta \omega })  \label{G1} \\
G_{2}({\ \omega }) &=&-1/\pi \mathrm{Im}\langle \langle \sigma
_{z}^{B}|\,\sigma _{z}^{B}\rangle \rangle _{\omega +i0^{+}}^{\prime
}/(1+e^{\beta \omega })  \label{G2}
\end{eqnarray}%
which lead to
\begin{equation}
J_{eff}^{TL}({\omega })=\frac{g_{0}^{2}\gamma (\left\vert \,\omega \right\vert )/\pi }{\left[
\left\vert \,\omega \right\vert -\eta \Delta _{B}-R(\left\vert \,\omega
\right\vert )\right] ^{2}+\gamma ^{2}(\left\vert \,\omega \right\vert )},
\label{Jeff_TL}
\end{equation}%
where the superscript "$TL$" indicate that it is the TL case (1st case), and this notation together with the superscript "$HO$" will be used in the following
when the distinguish between these two cases are needed. $\left\vert \,\cdots \right\vert $ means the absolute value, $R(\omega )$
and $\gamma (\omega )$ are the real and imaginary parts of $%
\sum_{k}V_{k}^{2}(2n_{k}+1)/(\omega -i0^{+}-\omega _{k})$,
\begin{eqnarray}
R(\omega ) &=&\wp \int_{0}^{\infty }\,d\omega ^{\prime }\frac{(\eta \Delta
_{B})^{2}J(\omega ^{\prime })\coth (\beta \omega ^{\prime }/2)}{(\omega
^{\prime }+\eta \Delta _{B})^{2}(\omega -\omega ^{\prime })},  \label{R} \\
\gamma (\omega ) &=&\pi (\eta \Delta _{B})^{2}J(\omega )\coth (\beta \omega
/2)/(\omega +\eta \Delta _{B})^{2},  \label{r}
\end{eqnarray}%
where $\wp $ means the Cauchy principal value.

To check the result, one can assume the pole of $G_{1}({\omega })$ and $G_{2}({\omega })$ to be $\omega
_{B}\pm \gamma (\omega _{B})$, where $\omega _{B}$ is the solution of
equation: $\omega -\eta \Delta _{B}-R(\omega )=0$. And Eq.~(\ref{G1}) and ( %
\ref{G2}) can be evaluated by using residue theorem,
\begin{eqnarray}
G_{1}({t}) &=&e^{-\gamma (\omega _{B})t}\left( e^{i{\omega }%
_{B}t}n_{B}^{\uparrow }+e^{-i\omega _{B}t}n_{B}^{\downarrow }\right) ,
\label{G1approx} \\
G_{2}({t}) &=&e^{-\gamma (\omega _{B})t}\left( e^{-i{\omega }%
_{B}t}n_{B}^{\uparrow }+e^{i\omega _{B}t}n_{B}^{\downarrow }\right) .
\label{G2approx}
\end{eqnarray}%
where, $n_{B}^{\uparrow }=1-n_{B}^{\downarrow }=\frac{1}{e^{\beta \omega
_{B}+1}}=\frac{e^{-\omega _{B}/2T}}{2\cosh (\omega _{B}/2T)}$. These
two-time correlation functions have the similar forms with those obtained in
the Born-Markov approximation \cite{Montina2008}. The difference is that
the tunneling frequency $\Delta _{B}$ is replaced by the renormalized
frequency $\omega _{B}$.

\subsection{2nd Case: DHO environment}

In the bosonic case, the equation of motion reads
\begin{equation}
\omega \langle \langle A|\,B\rangle \rangle _{\omega }=\langle [A,B]\rangle +\langle \langle [A,H_{B}]|\,B\rangle \rangle _{\omega },
\end{equation}%
where $\langle \langle A|\,B\rangle \rangle $ represents the Fourier
transform of the Green's function $-i\theta (t)\langle [A,B]\rangle
_{\beta }$. And the corresponding FDT is expressed as
\begin{eqnarray}
G_{1}({\omega }) &=&-1/\pi \mathrm{Im}\langle {\langle }Q^{B}|\,Q^{B}\rangle
\rangle _{\omega +i0^{+}}/(1-e^{-\beta \omega }), \\
G_{2}({\ \omega }) &=&-1/\pi \mathrm{Im}\langle {\langle }%
Q^{B}|\,Q^{B}\rangle \rangle _{\omega +i0^{+}}/(e^{\beta \omega }-1),
\end{eqnarray}%
where $Q^{B}=B^{\dagger }+B$ in this case. Since the DHO model is exactly solvable \cite{Costa2000}, the equation chain is
self-closed automatically without any approximation. Similarly, one get
\begin{equation}
J_{eff}^{HO}({\omega })=\frac{4g_{0}^{2}{\Omega }^{2}J(\omega )\coth (\beta \omega /2)}{\left[
\omega ^{2}-{\Omega }^{2}-2{\Omega H}_{R}(\omega )\right] ^{2}+\left[ 2\pi {%
\Omega }J(\omega )\right] ^{2}}  \label{Jeff_HO}
\end{equation}%
with%
\begin{equation}
{H}_{R}(\omega )=\wp \int_{0}^{\infty }dx\frac{J(x)}{\omega -{x}}%
-\int_{0}^{\infty }dx\frac{J(x)}{\omega +{x}}+2\int_{0}^{\infty }dx\frac{J(x)%
}{{x}}.  \label{HR_HO}
\end{equation}

\section{Results and discussions}

\subsection{1st case: TLF environment}

Here we would like to summarize the approximations we have made in this case.
Three approximations are made: The first one is the 2nd order approximation to the TL-bath
coupling in obtaining the transformed Hamiltonian. The second one is to approximate $%
(b_{k}^{\dag }+b_{k})(b_{k^{\prime }}^{\dag }+b_{k^{\prime }})\approx
(2n_{k}+1)\delta _{kk^{\prime }}$ to make the equation chain of motion self-closed. The third one
is the Born approximation in deriving the master equation (\ref{ME_S}).
Therefore, our treatment is applicable for $\alpha <1.0$ (see Eq.~(\ref{J_TL}%
)) and $g_{0}{\ll }\Delta _{A},\Delta _{B}$.

In order to calculate $P(t)$, we have to specify the spectral density $%
J^{TL}(\omega )$. We will use the piezoelectric spectral density, which describes the
decoherence of a double quantum dots (DQD) qubit manufactured with GaAs \cite%
{Brandes1999,Wu2005},
\begin{equation}
J^{TL}(\omega )=\alpha \omega \left( 1-\frac{\omega _{d}}{\omega }\sin \frac{%
\omega }{\omega _{d}}\right) e^{-\omega ^{2}/2\omega _{l}^{2}},  \label{J_TL}
\end{equation}%
where $\omega_d$ is related to the center to center distance, and $\omega_l$ to the dot size.
Typically, $\omega_d\sim0.01(ps)^{-1}$ and $\omega_l\sim1(ps)^{-1}$ \cite%
{Wu2005}. In the limit of $\omega_d\rightarrow0$, one can find that, Eq.~(\ref{J_TL}) goes back to the widely used Ohmic spectrum \cite{Weiss1999}.

Now, we are in position of calculating $P(t)$. We first use Eq.~(\ref{xi}), (%
\ref{eta}) and (\ref{J_TL}) to get $\eta $ self consistently, then calculate $%
P(t)$ numerically according to Eq.~(\ref{Ps}), (\ref{Fs}), (\ref{Jeff_TL})-(\ref{r}) and (\ref{J_TL}). $P(t)$ are reported as a function of time in the main plots of Fig.~2
and Fig.~3 for $\Delta _{A}=\Delta _{B}=0.1\omega _{l}$, $g_{0}=0.1\Delta
_{A}$ and $\omega _{d}=0.05\omega _{l}$. In Fig.~2, where $\alpha =0.3$, it shows simple oscillation for the qubit dynamics and  the
decoherence is reduced by increasing the bath temperature $T$. However, in
Fig.~3, where $\alpha =0.01$, the beating dynamics appears, and the coherence is not meliorated but rather
damaged with increasing $T$. In Fig.~4, we set fixed detuning
between A-B: $\delta =\Delta _{B}-\Delta _{A}=0.1\omega _{l}$ and $%
g_{0}=0.1\Delta _{A}$, the coherence is meliorated with increasing $T$ in
Fig.~4(a) where $\alpha =0.3$, but damaged in Fig.~4(b) where $\alpha =0.1$. And the beating is totally suppressed for the off-resonance cases.

If we consider the Weisskopf-Wigner approximation, the damping rate
of the qubit is $\pi J_{eff}(\omega)$ \cite{Weiss1999}. One can find the above
results are in consistency with the property of the effective spectral density.
The main difference of the two cases is that the
temperature dependent term $\coth (\beta \omega /2)$ enters into both the
numerator and the denominator in the 1st case (See Eq.~(\ref{Jeff_TL}) and Eq.~(\ref{r})), whereas it
only appears in the numerator in the 2nd case (see Eq.~(\ref{Jeff_HO})). This difference causes the
different qubit behaviors. In the following, we explore the properties of
the effective spectral density in detail.

The effective spectral density $J_{eff}^{TL}(\omega )$ are reported in the insets of
Fig.~2 and Fig.~3 according to Eq.~(\ref{Jeff}). One can find that they are
consistent with the numerical results given in the main plots. We should note that, in Fig.~3, level repulsion occurs, two
characteristic frequencies $\omega _{B}\approx \Delta _{A}\pm g_{0}$ are
dominating the qubit dynamics (see the inset (a) of Fig.~3). Therefore, in
Fig.~3(b), it is $J_{eff}(\Delta _{A}\pm g_{0})$, rather than $J_{eff}(\Delta _{A})$, that determines the damping
rate of the qubit. And we can find the damping rate is indeed enhanced with
bath temperature, which explains the results in the main plot of Fig.~3.
Note that, the frequency shift (from $\Delta _{A}$ to $\Delta _{A}\pm g_{0}$) plays an important role here. Thus, this
effect can not be predicted by using Lindblad formula, where the frequency
shift is not considered \cite{Montina2008}.

\subsection{2nd Case: DHO environment}

Since the DHO is exactly solvable \cite{Costa2000}, the only
approximation is the weak-coupling approximation in deriving the master
equation. Here we conveniently use the Ohmic spectral density in the Drude
form,%
\begin{equation}
J^{HO}(\omega )=\frac{\Gamma \omega }{1+\omega ^{2}/\omega _{c}^{2}},
\label{J_HO}
\end{equation}%
so that the integral in (\ref{HR_HO}) can be evaluated explicitly \cite%
{Costa2000}. In the limit $\omega _{c}\rightarrow \infty$, according to Eq.~(\ref%
{Jeff_HO}) and (\ref{HR_HO}), we obtain
\begin{equation}
J_{eff}^{HO}({\omega })=\frac{4g_{0}^{2}{\Gamma \omega \Omega }^{2}\coth
(\beta \omega /2)}{\left( \omega ^{2}-{\Omega }^{2}\right) ^{2}+\left( 2\pi {%
\Gamma \Omega \omega }\right) ^{2}}.  \label{Jeff_HO2}
\end{equation}%
One can check that this effective spectral density is in consistency with
the previous studies, where the mapping between the model and the SBM are
considered \cite{Garg1985,Tian2002}.

According to Eq.~(\ref{Ps}), (\ref{Fs}) and (\ref{Jeff_HO2}), $P(t)$ is
obtained numerically. Here we report $P(t)$ as a function of time in the main plot of Fig.~5
and Fig.~6 for $\Delta _{A}=\Omega $, $g_{0}=0.1\Delta _{A}$. In Fig.~5,
where $\Gamma =0.3$, the decoherence is enhanced by increasing the bath
temperature $T$, which is different compared to the first case (see Fig.~2). In Fig.~6,
where $\Gamma =0.01$, the coherence is also not meliorated with increasing $T
$. In Fig.~7, we set fixed detuning
between A-B: $\Omega=2\Delta _{A}$ and $%
g_{0}=0.1\Delta _{A}$, the decoherence is enhanced with increasing $T$ both in
Fig.~7(a) where $\Gamma=0.3$, and in Fig.~7(b) where $\Gamma =0.01$. In the 2nd case, we can not find a parameter set, in which the coherence is meliorated with increasing $T$. From the results, one can also find that the qubit shows beating dynamics when $\Gamma$ is small for the on-resonance case, and it is suppressed for large $\Gamma$ or finite detuning.

The effective spectral density $J_{eff}^{HO}(\omega )$ explains the results
given above. It is reported in the insets of Fig.~5 and Fig.~6 according to Eq.~(\ref%
{Jeff_HO}). In Fig.~5(b), $J_{eff}(\omega )$ increases with temperature at $%
\omega \approx \Delta _{A}$, rather than decreases as appeared in Fig.~2(b).
In Fig.~6(b), similarly to the 1st case, two characteristic frequencies
appear (see the inset (a) of Fig.~6). However, we should note that, the splitting
is increasing significantly with increasing temperature, which is also
affecting the temperature dependent properties. The dominant frequencies of $%
P(t)$ are marked in Fig.~6(b). Attention should pay to the frequencies with smaller damping, which determines the long time dynamics. We can see
the corresponding $J_{eff}(\omega )$ of the smaller values are of the same order, which explains why
the damping rate of $P(t)$ in the main plot is almost not changing with different
temperatures.

\subsection{Conclusion}

In conclusion, we compared the non-Markov dynamics of a qubit under the decoherence
of two kinds of reservoirs. Within the
weak-coupling approximation, the qubit dynamics boils down to the correlation
functions of the structured bath.
In the 1st case, where the problem is not exactly solvable, we
obtain the correlation functions by the second order perturbation based on a unitary transform. We find that, the decoherence of the qubit can be reduced with increasing bath temperature $T$ when $g_0<\alpha$, which agree with the results in Ref. \cite{Montina2008}. Whereas, it can only be enhanced when $g_0>\alpha$. The results are
checked in both on-resonance and off-resonance cases. In the 2nd case as a comparison, where the correlation functions are exactly solvable, we find that the decoherence of the qubit can not meliorated with
increasing $T$. In both cases, the qubit shows beating dynamics when TL-boson (or HO-boson) coupling is small for the on-resonance case, and it is suppressed for large coupling or finite detuning.

We thank Z. L\"{u}, X. Cao for discussions with P. Huang early in this work
and S. Ashhab for comments and discussion. This work was supported by the National Natural Science Foundation
of China (Grant No.10734020) and the National Basic Research Program of China (Grant No. 2011CB922202).

\appendix

%
%

\section*{Figures Captions}

{Fig.~1:} {Diagrammatic sketch of a qubit coupled with structured environments. The environment in the 1st case consists of a two level system coupled to a bath. The environment in 2nd case is a damped harmonic oscillator.
}

{Fig.~2:} (1st case) {$P(t)$ as a function of time for the on-resonance case ($%
\Delta_A=\Delta_B$), where the decoherence is reduced with $T$. Inset (a):
Fourier analysis of the main plot. Inset (b): The effective spectral density $J_{eff}(\omega )$. $\pi J_{eff}(\Delta _{A})$ indicates the
damping rate $\gamma _{A}$. One can see that $\gamma _{A}$ is reduced with
increasing $T$, which is consistent with the main plot.}

{Fig.~3:} (1st case) {$P(t)$ as a function of time for the on-resonance case ($%
\Delta_A=\Delta_B$), where the decoherence is enhanced with $T$. Inset (a): Fourier analysis of $P(t)$. One can see that two frequencies are dominating the dynamics and the peaks locate at $\Delta _{A}\pm {g_{0}}$. Inset (b): The effective spectral density $J_{eff}(\omega )$. Here, it is not ${\pi}J_{eff}(\Delta _{A})$ but $\pi J_{eff}({\Delta _{A}\pm {g_{0}}})$ indicates the damping rate $\gamma _{A}$.

{Fig.~4:} (1st case) {$P(t)$ as a function of time for the off-resonance case ($%
\Delta_A\neq\Delta_B$). (a): The decoherence is enhanced with $T$. (b): The decoherence is reduced with $T$.

{Fig.~5:} (2nd case) {$P(t)$ as a function of time, where the decoherence is enhanced with $T$. Inset (a): Fourier analysis of the main plot. Inset (b): The effective spectral density $J_{eff}(\omega )$. $\pi J_{eff}(\Delta _{A})$ indicates the
damping rate $\gamma _{A}$. One can see that $\gamma _{A}$ is enhanced with
increasing $T$, which is consistent with the main plot.

{Fig.~6:} (2nd case) {$P(t)$ as a function of time, where the decoherence is enhanced with $T$. Inset (a): Fourier analysis of the main plot.
One sees that two frequencies are dominating the dynamics and the splitting of the peaks increases with temperature. Inset (b): The effective spectral density $J_{eff}(\omega )$. The square, triangle and circle points correspond to the dominant frequencies of $P(t)$ in different temperature respectively. One can see that smaller $J_{eff}$'s, which characterize long time dynamics, are almost the same for three different temperatures. This is the reason why the damping rate of $P(t)$ is almost not changing with different temperatures.}

{Fig.~7:} (2st case) {$P(t)$ as a function of time for the off-resonance case ($%
\Delta_A\neq\Omega$). (a): $\Gamma=0.3$. (b): $\Gamma=0.01$. The decoherence is enhanced with $T$ in both cases.

\clearpage
\begin{figure}[tbp]
\includegraphics[scale=0.8]{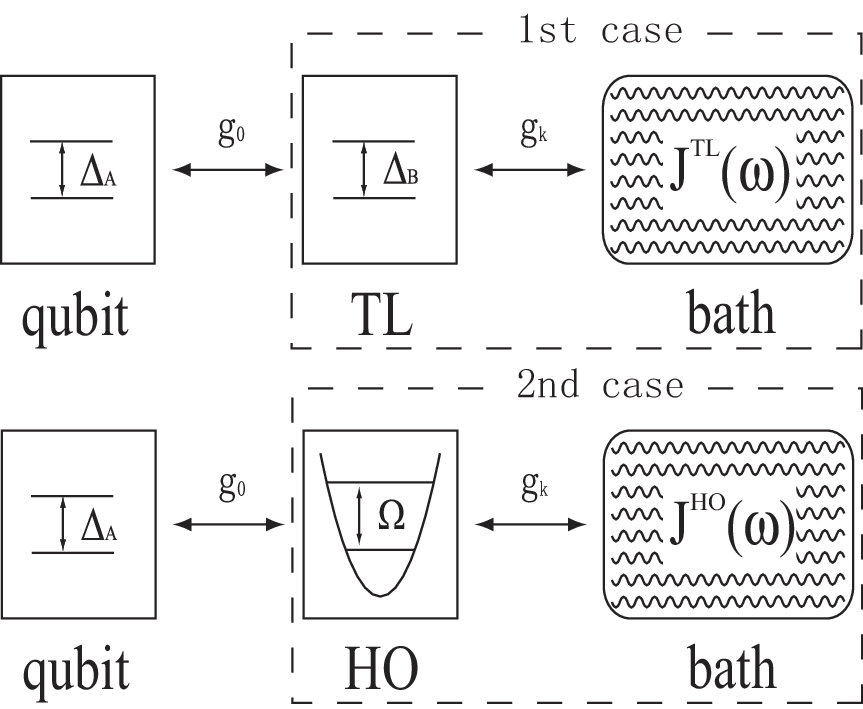}
\caption{Fig.~1}
\end{figure}
\clearpage
\begin{figure}[tbp]
\includegraphics[scale=0.6]{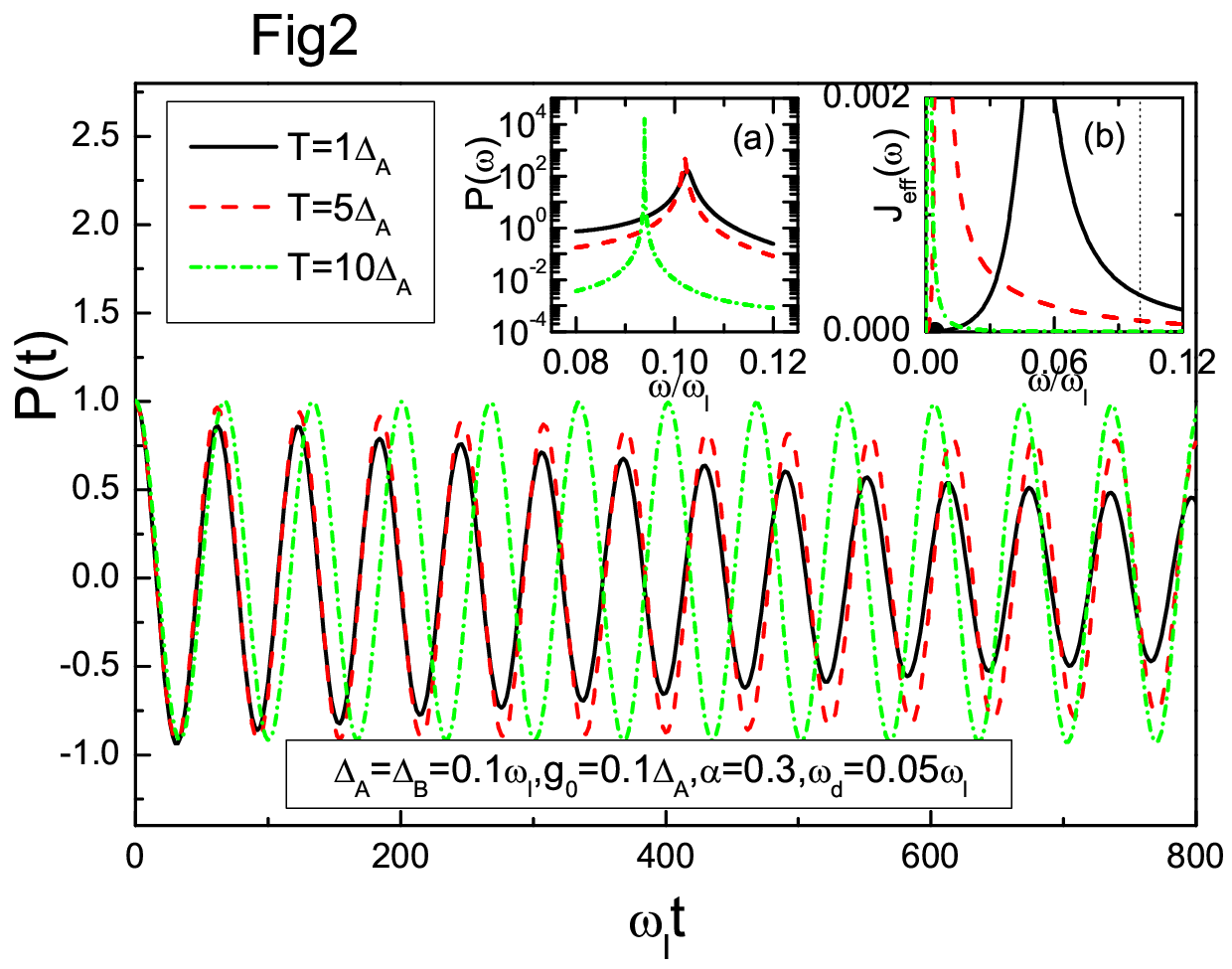}
\caption{Fig.~2}
\end{figure}
\clearpage
\begin{figure}[tbp]
\includegraphics[scale=0.6]{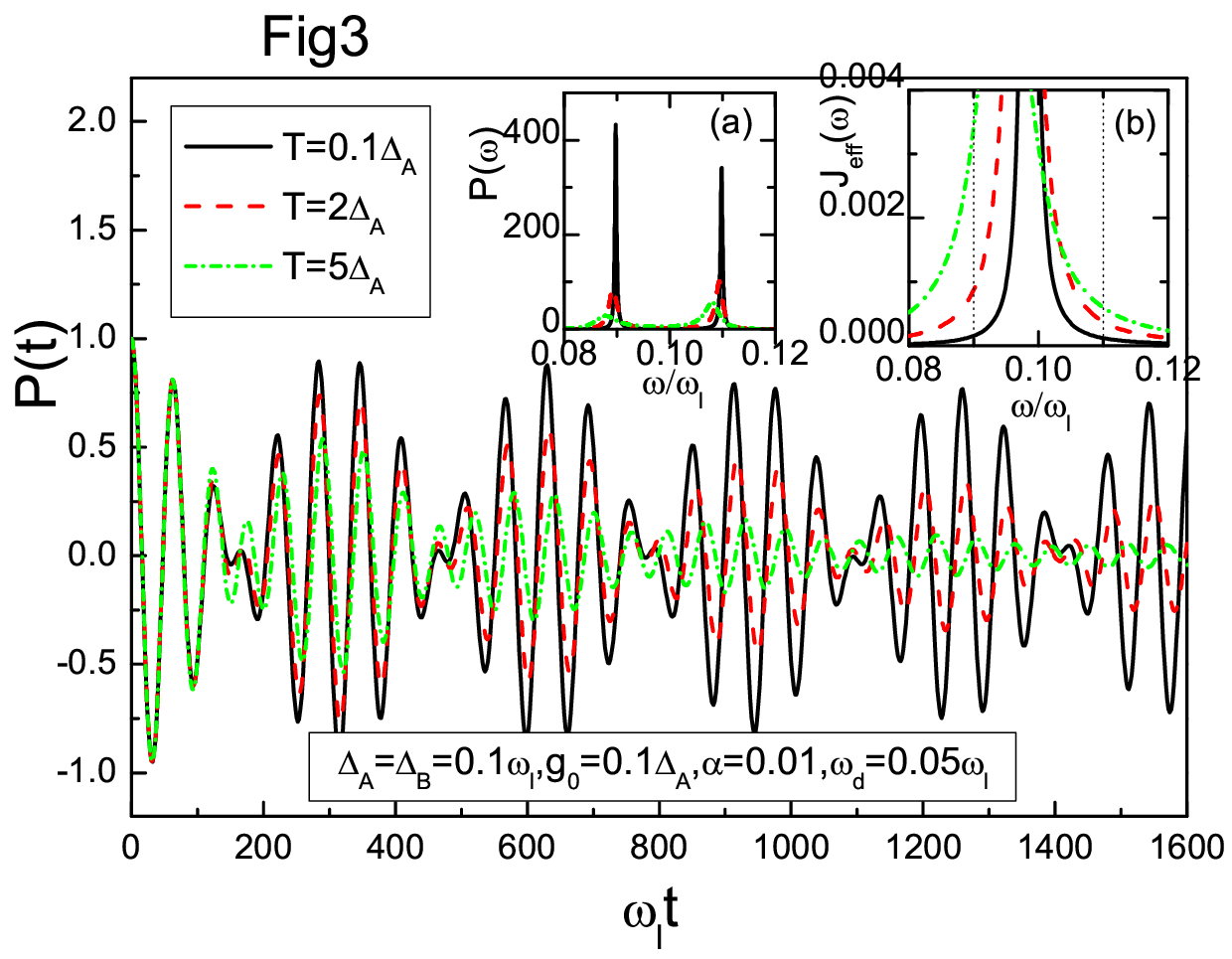}
\caption{Fig.~3}
\end{figure}
\clearpage
\begin{figure}[tbp]
\includegraphics[scale=0.6]{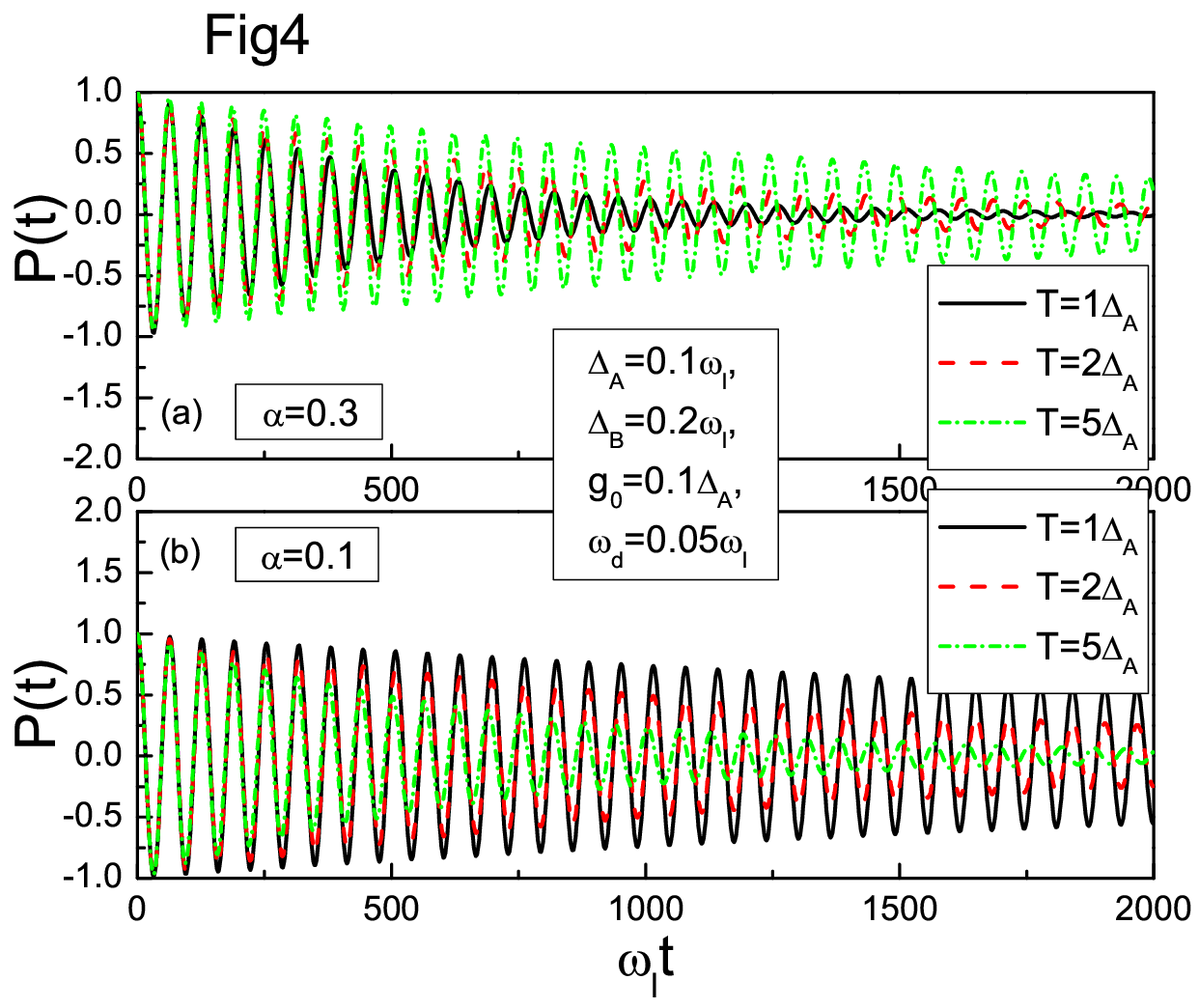}
\caption{Fig.~4}
\end{figure}
\clearpage
\begin{figure}[tbp]
\includegraphics[scale=0.6]{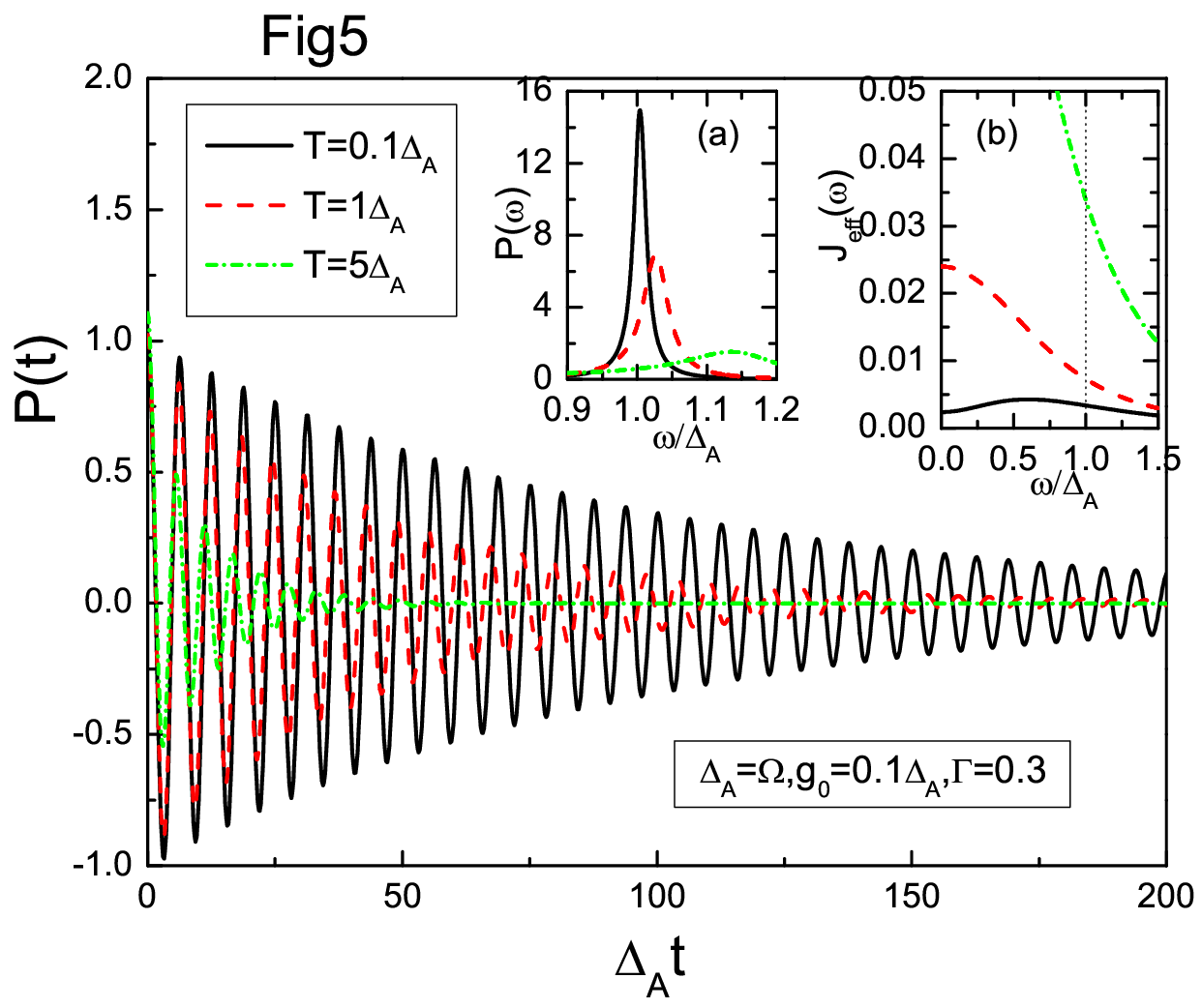}
\caption{Fig.~5}
\end{figure}
\clearpage
\begin{figure}[tbp]
\includegraphics[scale=0.6]{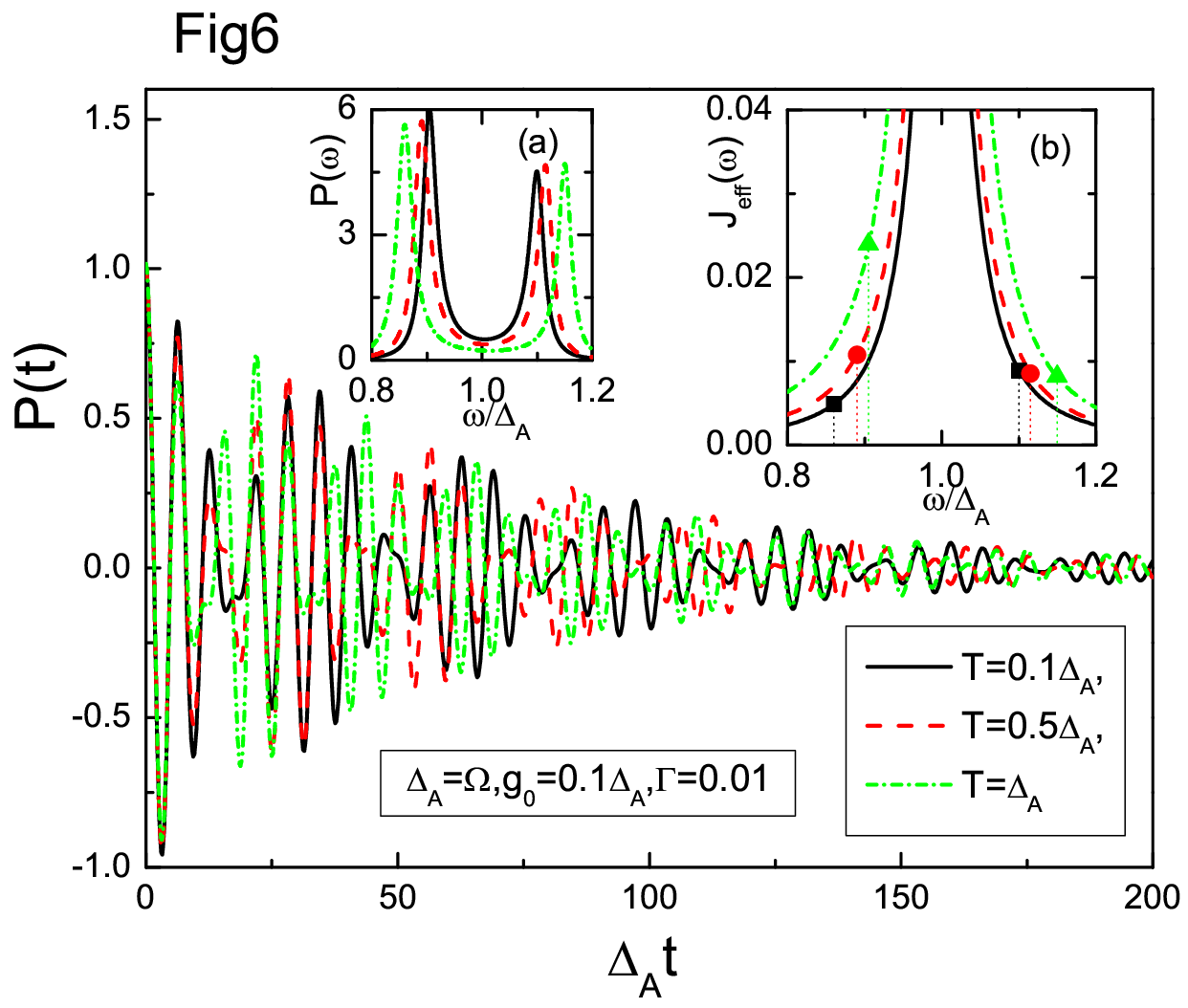}
\caption{Fig.~6}
\end{figure}
\clearpage
\begin{figure}[tbp]
\includegraphics[scale=0.6]{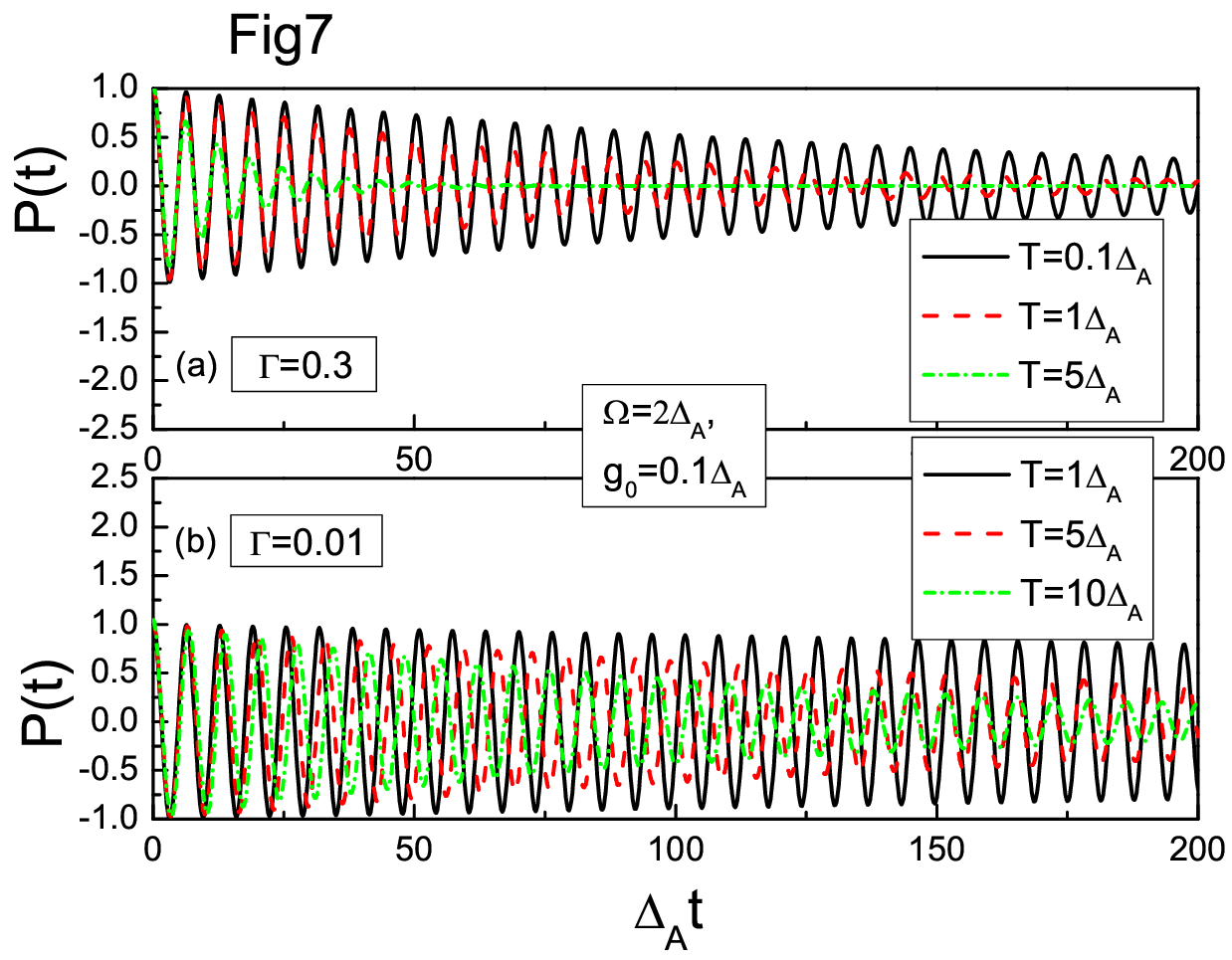}
\caption{Fig.~7}
\end{figure}

\end{document}